\documentclass[%
reprint,
showpacs,showkeys
 amsmath,amssymb,
 aps,
]{revtex4-1}

\usepackage{epsfig}
\usepackage{graphicx}
\usepackage{dcolumn}
\usepackage{bm}

\usepackage{hyperref}

\begin{document}

\title{First bolometric measurement of the two neutrino double beta decay of $^{100}$Mo with a ZnMoO$_4$ crystals array\\}

\author{L.~Cardani$^{1,~2}$}
\author{L.~Gironi$^{3,~4}$}
\email{Corresponding author: luca.gironi@mib.infn.it}
\author{N.~Ferreiro Iachellini$^{4}$}
\author{L.~Pattavina$^{5}$}
\author{J.W.~Beeman$^{6}$}
\author{F.~Bellini$^{1,~2}$}
\author{N.~Casali$^{5}$}
\author{O.~Cremonesi$^{4}$}
\author{I.~Dafinei$^{2}$}
\author{S.~Di Domizio$^{7}$}
\author{F.~Ferroni$^{1,~2}$}
\author{E.~Galashov$^{8}$}
\author{C.~Gotti$^{4}$}
\author{S.~Nagorny$^{5}$}
\author{F.~Orio$^{2}$}
\author{G.~Pessina$^{4}$}
\author{G.~Piperno$^{1,~2}$}
\author{S.~Pirro$^{4}$}
\author{E.~Previtali$^{4}$}
\author{C.~Rusconi$^{4}$}
\author{C.~Tomei$^{2}$}
\author{M.~Vignati$^{2}$}

\affiliation{$^{1}$Dipartimento di Fisica, Sapienza Universit\`a di Roma, Roma I-00185, Italy}
\affiliation{$^{2}$INFN- Sezione di Roma, Roma I-00185, Italy}
\affiliation{$^{3}$Dipartimento di Fisica, Universit\`a di Milano-Bicocca, Milano I-20126 - Italy}
\affiliation{$^{4}$INFN - Sezione di Milano Bicocca, Milano I-20126 - Italy}
\affiliation{$^{5}$INFN - Laboratori Nazionali del Gran Sasso, Assergi (L'Aquila) I-67010 - Italy}
\affiliation{$^{6}$Lawrence Berkeley National Laboratory, Berkeley, California 94720 - USA}
\affiliation{$^{7}$Dipartimento di Fisica, Universit\`a di Genova and INFN - Sezione di Genova, I-16146 Genova - Italy}
\affiliation{$^{8}$Department of Applied Physics, Novosibirsk State University, Novosibirsk 630090 - Russia}

\date{\today}

\begin{abstract}

The large statistics collected during the operation of a ZnMoO$_4$ array, for a total exposure of 1.3 kg $\cdot$ day of $^{100}$Mo, allowed the first bolometric observation of the two neutrino double beta decay of $^{100}$Mo. The observed spectrum of each crystal was reconstructed taking into account the different background contributions due to environmental radioactivity and internal contamination. The analysis of coincidences between the crystals allowed the assignment of constraints to the intensity of the different background sources, resulting in a reconstruction of the measured spectrum down to an energy of $\sim$300 keV. The half-life extracted from the data is T$_{1/2}^{2\nu}$= [7.15 $\pm$ 0.37 (stat) $\pm$ 0.66 (syst)] $\cdot$ 10$^{18}$ y.

\end{abstract}

\maketitle

\section{INTRODUCTION\\}

The two-neutrino double beta decay (2$\nu$DBD) of atomic nuclei,

\begin{displaymath}
(A,Z) \rightarrow (A,Z+2) + 2e^-+2\overline{\nu}_e
\end{displaymath}

is the rarest nuclear weak process experimentally observed. Being a second order process in the Standard Model (SM), characterized by an extremely low decay rate, the 2$\nu$DBD is observable for even-even nuclei whose single $\beta$ decay is forbidden or hindered by large spin difference. This decay has been detected so far for twelve nuclides, with half-lives ranging from 10$^{18}$ to 10$^{24}$ yr \cite{Barabash, Tretyak}. 

A huge experimental effort of the Particle Physics community is now focused on the detection of the neutrinoless double beta decay (0$\nu$DBD), the SM forbidden counterpart of  2$\nu$DBD, that violates lepton number by two units and whose observation would prove that neutrinos have a Majorana nature.

The measurement of the half-life of the 2$\nu$DBD (T$_{1/2}^{2\nu}$) has various interesting theoretical implications that are crucial for 0$\nu$DBD search.
As suggested by~\cite{Rodin} and adopted in many subsequent works, data on half-lives of  2$\nu$DBD are used to fix the value of the strength parameter $g_{pp}$ involved in the calculation of the Nuclear Matrix Element (NME) of 0$\nu$DBD. These calculations require accurate nuclear models and suffer from large uncertainties.
In this respect, testing the model predictions on the 2$\nu$DBD plays a very important role. Since the 2$\nu$DBD mode connects the same initial and final nuclear ground states as the 0$\nu$DBD mode, the understanding of the nuclear structures embedded in the 2$\nu$DBD NME is an important step toward a reliable estimate of the 0$\nu$DBD NME. Charge-exchange reactions are one of the major tools to study the wave functions of nuclear models~\cite{*[{}][{ and references
therein.}] ChargeEx}. Comparing the values of the 2$\nu$DBD NME that can be derived from experimentally measured half-lives with the calculations based on charge-exchange experiments can help in constraining the parameter space of different nuclear models.

Among the isotopes that undergo double beta decay, $^{100}$Mo stands out not only for its large Q-value (3034 keV~\cite{Qvalue}) and quite high natural abundance (9.7\%~\cite{IsoAbund}), but also because of the relatively short half-life of the 2$\nu$DBD. The most accurate measurement of this transition has been performed by the NEMO3 collaboration, that set a value of T$_{1/2}^{2\nu}$ = [7.11 $\pm$ 0.02(stat) $\pm$ 0.54(syst)] $\cdot$ 10$^{18}$ y for the half-life of this decay~\cite{NEMO3}.
The 0$\nu$DBD has never been observed and the current best limit on the half-life of such a process is also set by NEMO3 as T$_{1/2}^{0\nu} >$ 1.1 $\cdot$ 10$^{24}$ y at 90\% C.L.~\cite{NEMO3_0nDBD}.

The excellent result of NEMO3 on the 2$\nu$DBD of $^{100}$Mo relies on the high statistics and the low background. This was achieved by means of the 3D reconstruction of the topology of the events provided by the tracking chamber, combined with the calorimetric information. However, because of the detector low energy resolution (FWHM$>$10\%) and the short half-life of the 2$\nu$DBD, the sensitivity on the 0$\nu$DBD is limited because of the unavoidable background due to the 2$\nu$DBD. The possibility to perform a measurement with an array of detectors with a high energy resolution, such as bolometers, is indeed very appealing~\cite{ZnMoO4_array}.

In this work we present the first bolometric measurement of the 2$\nu$DBD half-life of $^{100}$Mo, performed with an array of three natural ZnMoO$_4$ cryogenic bolometers, for a total mass of  811 g, at the Gran Sasso National Laboratory (LNGS) of INFN.

\section{EXPERIMENTAL TECHNIQUE}

Bolometers are very sensitive calorimeters operated at cryogenic temperatures. In this experiment, bolometers measure the energy deposited by an interacting particle through a corresponding temperature rise. These solid-state detectors share with Ge diodes the capability of achieving excellent energy resolution ($\sim$5 keV FWHM from several keV to several MeV) with sizable active mass devices. Our bolometers apply the calorimetric (source = detector) approach to the detection of rare decays: the source isotope is part of the active mass of the detector. The latter consists of two elements: a single crystal that plays the role of the calorimetric mass, and a sensor that measures the amount of energy converted into heat in the crystal, transducing the phonon signal into an electrical one. 

Zync molibdate (ZnMoO$_4$) crystals are suitable detectors for double beta decay studies. Following the promising results obtained in the first measurement with a small crystal~\cite{ZnMoO4_First}, recent studies~\cite{ZnMoO4_Large} have demonstrated the possibility to grow large crystals with high radio-purity and excellent energy resolution.

\subsection{Detectors and experimental setup}

The three ZnMoO$_4$ crystals used in this work were grown using the low-thermal-gradient Czochralski technique at the Nikolaev Institute of Inorganic Chemistry (NIIC, Novosibirsk, Russia) and at the Novosibirsk State University (NSU, Novosibirsk, Russia). The starting materials for the crystal growth were high purity ZnO (produced by Umicore) and MoO$_3$, synthesized by NIIC. An irregular shape of the crystals was chosen as a compromise between the largest achievable crystal size and the minimization of visible defects that could spoil the bolometric performances. The crystal masses were 247 g, 235 g and 329 g, respectively.

\begin{figure}[t]
\begin{center}
\includegraphics[width=.8\linewidth]{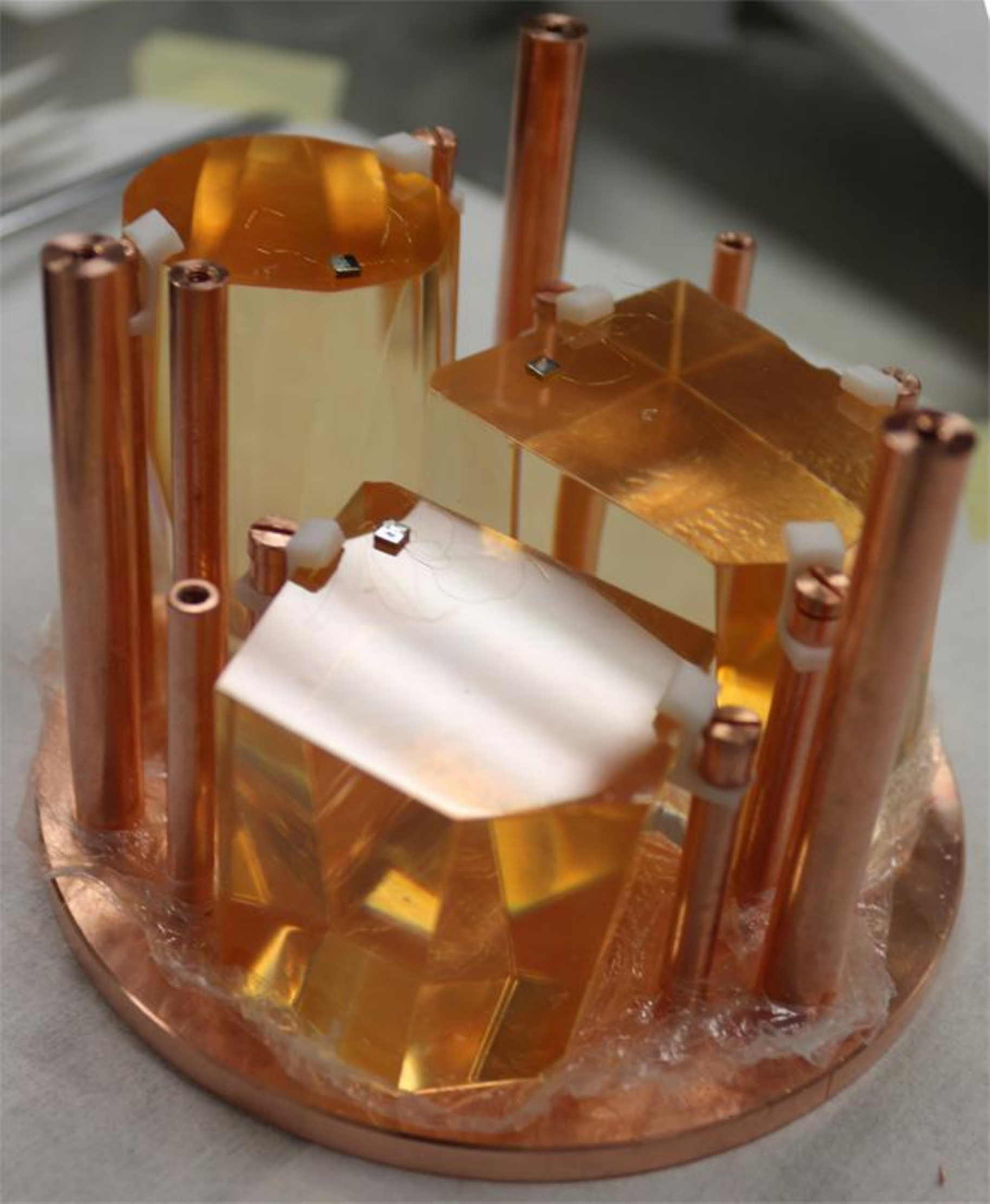}
\end{center}
\caption{The three ZnMoO$_4$ crystals mounted in the copper structure. The crystals are held in place by means of PTFE clamps. The thermistors glued to the top of the crystals are also visible.}
\label{fig:setup}
\end{figure}

The crystals were mounted in a mechanical structure (see Fig.~\ref{fig:setup}), made of copper elements and PTFE clamps and covered with several polyethylene layers in order to suppress the background produced by $\alpha$ contamination~\cite{CUORE_Surf}.  The bolometers were arranged in a closely packed setup in order to increase the geometrical efficiency for the coincidence analysis. Each detector was instrumented with a 3$\times$3$\times$1 mm$^3$ Neutron Trasmutation Doped (NTD) Ge thermistor, thermally coupled to the crystal via 6 epoxidic glue spots of $\sim 600 \mu$m diameter and $\sim 50 \mu$m height. The NTD is a resistive device made of semiconducting material, which converts temperature variations into resistance variations. When the thermistor is biased with a constant current, any resistance variation produces a voltage pulse, that constitutes the signal. 

The mechanical structure of Fig.~\ref{fig:setup} was enclosed in a copper shield. The entire setup, thermally coupled to the mixing chamber of a $^3$He/$^4$He dilution refrigerator installed in Hall C of LNGS, was surrounded by a ancient lead shield~\cite{Roman_Lead} with the purpose of reducing the $\gamma$ radioactivity induced by the cryostat shields and the laboratory environment. The cryostat was cooled down to a temperature of $\sim$10 mK.

The voltage signals produced by the NTD thermistors were amplified and filtered by front-end electronics and fed into the ADC. For each triggered event, the entire signal waveform was sampled, digitized and saved to disk. Pulse height and pulse-shape parameters were computed off-line by means of the Optimum Filter (OF) technique~\cite{Gatti}. The cryogenic facility, the electronic read-out and the data acquisition are described in detail in~\cite{Cryo_Details, Front-end}. 

We performed calibration runs for a total of 174 h with a $^{228}$Th source and 20 h with a $^{40}$K source. The calibration function was a second order polynomial with zero intercept.

Pulse shape cuts were applied to suppress pile-up or noisy or spurious events that can produce a broadening or a deformation of the peaks, spoiling the energy resolution. 
An algorithm that allows energy-dependent cuts was used in order to compensate for the broadening of the shape parameter distributions at lower energy. It allows to obtain constant pulse shape cuts efficiencies over the whole energy spectrum considered for the study of the 2$\nu$DBD.

Finally, a coincidence analysis was performed in order to select events that took place in a single crystal (anti-coincidence events) and events that released energy in two crystals simultaneously (coincidence events). The coincidence window was set to 25 ms. We observed and identified also triple coincidences but we did not include them in the 2$\nu$DBD analysis because of the low detection efficiency.

The event selection efficiency after the pulse shape and anti-coincidence cuts was evaluated for each crystal on the $^{228}$Th and $^{40}$K calibration lines. The results show that the event selection efficiency does not depend on the energy. However, due to the rather large spread in the obtained values, the error on this quantity was set equal to the spread itself, obtaining 82$\pm$2\%, 84$\pm$3\% and 82$\pm$3\% for the three crystals, respectively.

We collected $\sim$900 h of data for a total mass of 34.6 g of $^{100}$Mo, amounting to an exposure of 1.3 kg $\cdot$ day. 

\section{Data analysis and results}
\subsection{General performances of the detectors}

The three crystals of the array showed similar performance in terms of energy resolution. The FWHM resolution measured on the most intense $\gamma$ peaks in calibration and background runs ranged between $\sim$10 keV and $\sim$20 keV. The energy region of interest extends between a few hundred keV and 3100 keV, just above the Q value of the 2$\nu$DBD of $^{100}$Mo. In this region the background is dominated by $\beta$/$\gamma$ events from internal contamination of the crystals and environmental radioactivity.

The bolometric performance, internal contamination and background rejection capability of the 329 g crystal were already investigated in a dedicated measurement~\cite{ZnMoO4_Large}, where the crystal was coupled to a light detector and operated as a scintillating bolometer. Scintillating bolometers are a development of pure bolometers where, besides measuring the heat signal, one can detect the scintillation light produced by ionizing events. The light yield, as well as the different time development of pulses produced by $\alpha$ and $\beta$/$\gamma$ particles, constitute two semi-independent discrimination tools that can provide an almost complete rejection of the $\alpha$ background~\cite{Bol_scint, PSA,ZnMoO4_Large}. This is crucial for 0$\nu$DBD searches, where $\alpha$ surface contaminations remain the main source of background for bolometric experiments~\cite{MC_Bucci}. 

For the study of 2$\nu$DBD, the rejection of $\alpha$ background is not a critical issue. In this work no light detectors were faced to the crystals to collect the scintillation light, neither we could rely on pulse shape discrimination. This was due to the very low operating temperature of the cryostat. A cooler base temperature resulted in a slower development of the thermal pulses. Among the different parameters that affect the pulse shape, it is particularly significant the RC low-pass filter due to the parasitic capacitances C of the wires used to bias and read the thermistor. At lower temperatures the thermistor resistance R increases considerably with a consequent lowering of the cut-off frequency. The higher frequencies of the signal that contain most of the information relating to the particle nature are therefore cut more at these temperatures. This leads to a loss of sensitivity on the small pulse shape differences between alpha and beta/gamma events due to the  scintillation decay time~\cite{PSA_Theor}.

\subsection{Background sources}
An accurate evaluation of the 2$\nu$DBD half-life requires a good understanding of the background in the region of interest.

Using a GEANT4-based Monte Carlo simulation of the experimental setup (crystals, assembly, shielding), we evaluated the spectral shape of the various components that are expected to contribute to the experimental spectrum of the three crystals: 

\begin{itemize}
\item \textbf{$^{100}$Mo DBD:} The 2$\nu$DBD was generated uniformly in each crystal. The non-relativistic Primakoff-Rosen approximation for the Coulomb interaction between the nucleus and the outgoing electrons was used to model the two electron energy spectrum in the Monte Carlo simulation.

\item  \textbf{$^{65}$Zn:} This isotope (T$_{1/2}$ = 244.3 d, Q-value = 1351.9 keV) is produced by the activation of $^{64}$Zn and decays into $^{65}$Cu via EC (98.5\%) or $\beta^+$-decay (1.5\%). The decay on the excited state of Cu (50.2\%) results in the immediate emission of a 1115.6 keV $\gamma$-ray. $^{65}$Zn was generated as a uniform bulk contamination of the crystals.

\item \textbf{$^{210}$Pb - $^{206}$Pb:} These isotopes belong to the $^{238}$U decay chain and are present as natural contaminants in almost every material. We simulated a $^{210}$Pb contamination both in the bulk of the crystals and in the copper elements facing them.

\item \textbf{$^{40}$K:} This isotope decays into $^{40}$Ca via $\beta$-decay (89.3\%, Q-value=1311 keV) or EC (10.7\%, Q-value=1505 keV) to $^{40}$Ar. Its presence is due to the natural radioactivity of the laboratory environment, to contamination in the cryostat shields and in the crystals themselves. The decays of this isotope were simulated both in the bulk of the crystals and around the experimental setup.

\item \textbf{$^{208}$Tl:} This isotope belongs to the $^{232}$Th decay chain. Its presence is due to the natural radioactivity of the laboratory environment and to contamination in the cryostat shields. The decays were simulated as uniformly distributed around the experimental setup. 

\end{itemize}

In addition to the simulated background sources, we accounted for the presence of a flat background due to $\alpha$ particles in the region of interest. This flat background is mainly due to surface contamination in the materials facing the detectors~\cite{MC_Bucci} but also to $\alpha$ decays from internal contaminations of the crystals that happen near the surface and release only a fraction of their energy inside the detectors. The flat $\alpha$ background was evaluated in the energy region between 2700 keV and 4000 keV where the number of $\beta$/$\gamma$ events is negligible.

\subsection{Fit model and results}

\begin{figure*}[t]
\begin{center}
\includegraphics[width=1.\linewidth]{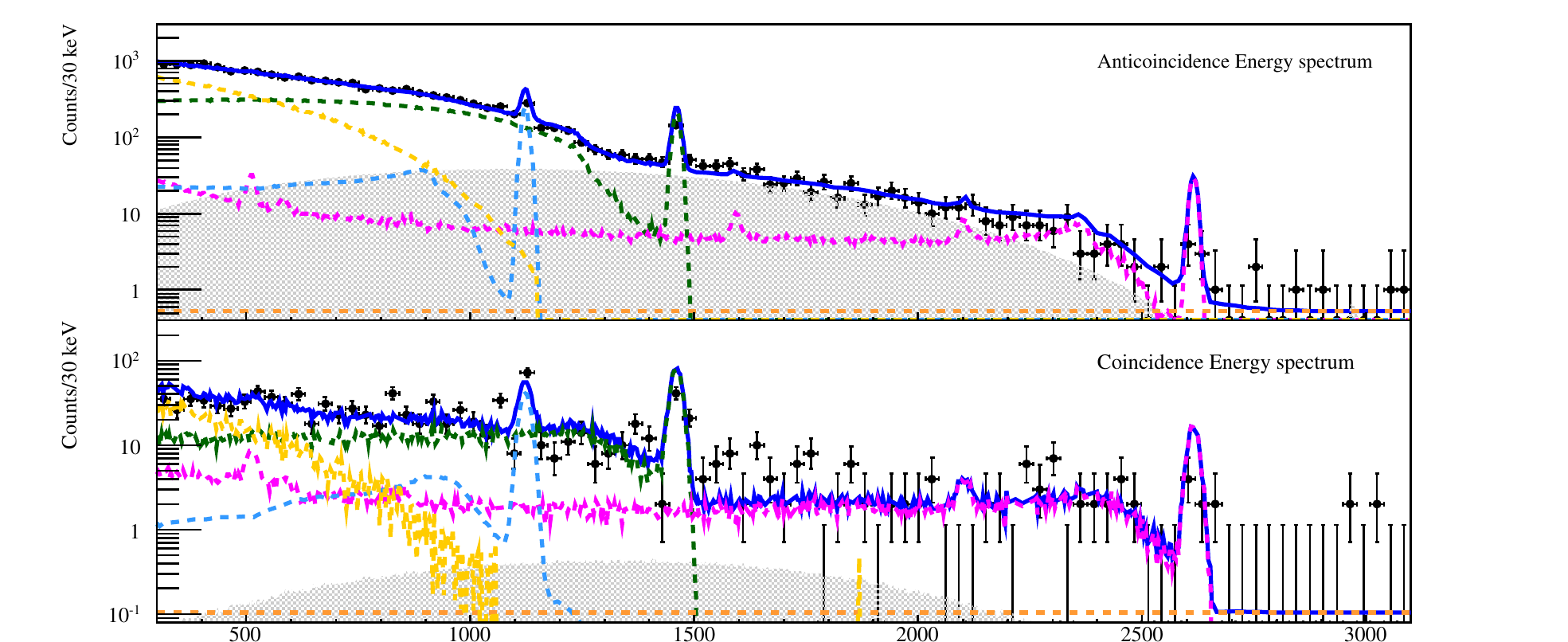}
\end{center}
\caption{Reconstruction of the $\sim$900 h background spectra measured with all the three ZnMoO$_4$ crystals. An energy threshold of 300 keV was set for all the detectors. The best fit line (solid blue) is shown. The components are 2$\nu$DBD (grey region), $^{65}$Zn (dotted light blue), internal and external $^{210}$Pb (dotted yellow), internal and external $^{40}$K (dotted green),  $^{208}$Tl (dotted pink), and flat $\alpha$ background (dotted orange).}
\label{fig:BackgroundFit}
\end{figure*}

The 2$\nu$DBD half-life is measured using a simultaneous, extended, unbinned maximum-likelihood fit~\cite{ROOFIT}. In a simultaneous fit, the data are divided into categories according to one or more discrete observables, and one or more fit parameters are shared among the categories. For the analysis presented in this work, the data are split according to the crystal and the multiplicity. This means we have six categories: the three anti-coincidence energy spectra of each crystal and the three coincidence total energy spectra (the total energy is the sum of the two coincident energy deposits) of each crystal. The six categories were fitted simultaneously with a function F constructed as a linear combination of the 2$\nu$DBD signal and the background sources listed above. The functional form of F can be written as follows:

\begin{eqnarray}
F(E,C,M) = \sum_i N_{i}(C,M) B_i (E,C,M) + \nonumber \\* N_{sig}(C,M) S (E,C,M) + B_{\alpha}(C,M) \nonumber
\end{eqnarray}

where the energy E ranges continuously from 300 keV to 3100 keV, and C, M are two discrete observables representing the crystal number (C = 1, 2, 3) and the multiplicity (M=1, 2). The index $i =1, ... , 6$ identifies the various background contributions to the measured spectrum, respectively {$^{65}$Zn, $^{210}$Pb (bulk and external), $^{40}$K (bulk and external) and $^{208}$Tl. The GEANT4 simulation was used to model the shape of the probability distribution functions  $B_i (E,C,M)$ (for the background sources) and $S (E,C,M)$ for the 2$\nu$DBD. The simulation is able to reproduce both the anti-coincidence and the coincidence spectrum of each crystal.
$N_{i}(C,M)$ is the number of events in the spectrum corresponding to the $i^{th}$ background source and $N_{sig}(C,M)$ is the number of signal events. Both depend on the crystal and the multiplicity, with the additional constraint that the ratio between the number of anti-coincidence and coincidence event is fixed by Monte Carlo.
This condition holds separately for each background source and for the signal. The quantity $B_{\alpha}(C,M)$ is a constant and represents the flat $\alpha$ background evaluated for each crystal and multiplicity.

Fixed parameters of the fit were the masses of the crystals and the live time of the measurement. The fit parameter shared among all the categories is the 2$\nu$DBD half-life $T_{1/2}^{2\nu}$.

Figure \ref{fig:BackgroundFit} shows experimental data together with the best fit model for the sum of the three detectors.

The best fit model has an expectation of 1643$\pm$49 events from the 2$\nu$DBD of $^{100}$Mo in the anti-coincidence spectrum and 16.0$\pm$0.5 events in the coincidence spectrum. The number of events for each crystal is reported in Table \ref{Table:nbb}. All this values refer to the number of events above thresholds ($\sim$300keV). The percentage of 2$\nu$DBD events below threshold has been estimated in $\sim$3\% through the analysis of the simulated spectrum and it is taken into consideration for the evaluation of the 2$\nu$DBD half-life.

\begin{table}[htb]
\centering
\caption{Number of events from the 2$\nu$DBD of $^{100}$Mo in the anti-coincidence spectrum and in the coincidence spectrum for each crystal.}
\label{Table:nbb}
\begin{tabular}{lcc}
\hline
Crystal mass ~~     & ~~ Anti-coincidence   ~~    & Coincidence	\\
\hline
\hline
247 g              &509 $\pm$ 26		& 4.4 $\pm$ 0.2 \\
\hline
235 g              &472 $\pm$ 24		& 5.4 $\pm$ 0.3\\ 
\hline
329 g              &661 $\pm$ 34		& 6.2 $\pm$ 0.3\\
\hline
\end{tabular}
\end{table}

\noindent
The corresponding half-life of the 2$\nu$DBD is

\begin{displaymath} 
T_{1/2}^{2\nu}(^{100}Mo)= [7.15 \pm 0.37 (stat) \pm 0.66 (syst)] \cdot 10^{18} y 
\end{displaymath}

We investigated various sources of systematic error. We evaluated the uncertainties related to the fit model by varying the energy threshold, the value of the flat $\alpha$ background and the order of interpolation of the simulated histograms used to model the shape of the signal and background components. The total contribution amounts to 2.2\%. The systematic error related to the use of Monte Carlo simulation was evaluated taking into account the uncertainty in the modeling of electromagnetic interactions in GEANT4~\cite{GEANT4_em} (5\%) and the inaccuracy in the description of the geometry of the experimental setup (estimated as an additional 5\%). The total contribution amounts to 7.1\%.
The uncertainty related to the modeling of the 2$\nu$DBD decay spectral shape was evaluated in~\cite{Kogler} for the $^{130}$Te double beta decay. In that case the use of the Primakoff-Rosen approximation caused an underestimation of the number of signal events of only a few per mil. Moreover, we made another test using the 2$\nu$DBD spectrum shape from~\cite{Iachello} and the fit result differs from the one obtained with the Primakoff-Rosen approximation only by 1.7\%.

Finally we included in the systematic error the uncertainties in the evaluation of the pulse shape cuts efficiency (4.6\%) and the uncertainty on the $^{100}$Mo isotopic abundance (3.1\%)~\cite{IsoAbund}. The combination in quadrature of all the contributions sums up to 9.3\% corresponding to $6.6 \cdot 10^{17}$ y.

The value of $T_{1/2}^{2\nu}$ measured with the ZnMoO$_4$ array is in perfect agreement with the one reported by the NEMO3 collaboration [7.11 $\pm$ 0.02(stat) $\pm$ 0.54(syst)] $\cdot$ 10$^{18}$ y.

\section{CONCLUSION} 

This work provides the first observation of the 2$\nu$DBD of $^{100}$Mo decay with an array of ZnMoO$_4$ crystals operating as bolometers.  The observed spectrum of each crystal was reconstructed taking into account the different background contributions due to environmental radioactivity and internal contamination. The analysis of coincidences between the crystals allowed to put additional constraints on the intensity of the different background sources. The measured half-life is T$_{1/2}^{2\nu}$= [7.15 $\pm$ 0.37 (stat) $\pm$ 0.66 (syst)] $\cdot$ 10$^{18}$ y, in agreement with the results of previous experiments.  

\begin{acknowledgments}
Part of the work was carried out thanks to LUCIFER Project, funded by the European Research  Council (FP7/2007-2013) grant agreement no. 247115. 
This work was also supported by the ISOTTA project, funded within the ASPERA 2nd Common Call for R\&D Activities. 

Thanks are due to F.Iachello and J.Kotila for fruitful discussions and for providing us precise numerical calculation of the electron distributions for the 2$\nu$DBD of $^{100}$Mo. We wish to express our gratitude to the LNGS mechanical workshop and in particular to E. Tatananni, A. Rotilio, A. Corsi, and B. Romualdi for continuous and constructive help in the overall set-up construction. Finally, we are especially grateful to M. Perego and M. Guetti for their invaluable help.
\end{acknowledgments}

\bibliography{ZnMoO4}

\end{document}